\documentclass{iaus}
\usepackage{graphicx}

\title[IAU 266.~~Binary Evolution and  Blue Stragglers] 
{Primordial Binary Evolution and  Blue Stragglers}

\author[Chen \& Han]   
{Xuefei Chen$^1$
 \and Zhanwen Han$^1$}

\affiliation{$^1$National Astronomical Observatories/Yunnan
Observatory, CAS, Kunming, 650011, P.R.China \\ email: {\tt
xuefeichen717@hotmail.com; zhanwenhan@hotmail.com}}

\pubyear{2009}
\volume{266}  
\pagerange{?}
 \jname{Star Clusters: Basic Galactic Building
Blocks Throughout Time And Space } \editors{Richard de Grijs and
Jacques Lepine}
\begin{document}

\maketitle

\begin{abstract}
Blue stragglers have been found in all populations. These objects
are important in both stellar evolution and stellar population
synthesis. Much evidence shows that blue stragglers are relevant
to primordial binaries. Here, we summarize the links of binary
evolution and blue stragglers, describe the characteristics of
blue stragglers from different binary evolutionary channels, and
show their consequences for binary population synthesis, such as
for the integrated spectral energy distribution, the
colour-magnitude diagram, the specific frequency, and the
influences on colours etc.. 
\end{abstract}

\firstsection 
\section{Introduction}
Blue stragglers (BSs) are an important population component in
stellar evolution as well as in star clusters. These objects have
remained on the main sequence for a time exceeding that expected
from standard stellar evolution theory, and they may affect the
integrated spectra of their host clusters by contributing excess
spectral energy in the blue and UV bands. Many mechanisms,
including single star models and binary models, have been
presented to account for the existence of BSs (see the review of
\cite{str93}). At present, it is widely believed that more than
one mechanism plays a role for the produce of BSs in one cluster
and that binaries are important or even dominant for the
production of BSs in open clusters and in the field
(\cite{lan07,dal08,sol08}). Binaries may produce BSs by way of
mass transfer, coalescence of the two components, binary-binary
collision and binary-single star collision. The collision of
binary-binary or binary-single may lead binaries to be tighter or
farther apart, and it is relevant to dynamics and environment in
the host cluster. In this contribution, we are only concerned with
BSs resulting from the evolutionary effect of primordial binaries,
i.e. mass transfer and coalescence of two components.

\section{Evolutionary channels}

\begin{figure}[t]
\begin{center}
 \includegraphics[width=1.7in,angle=270]{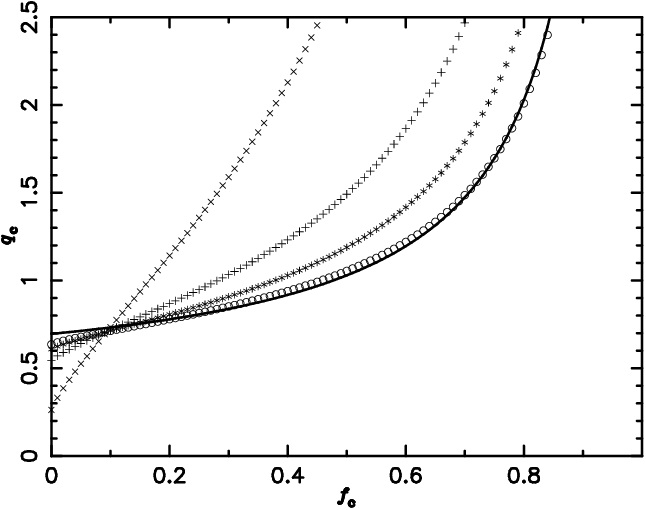}
 \includegraphics[width=1.8in,angle=270]{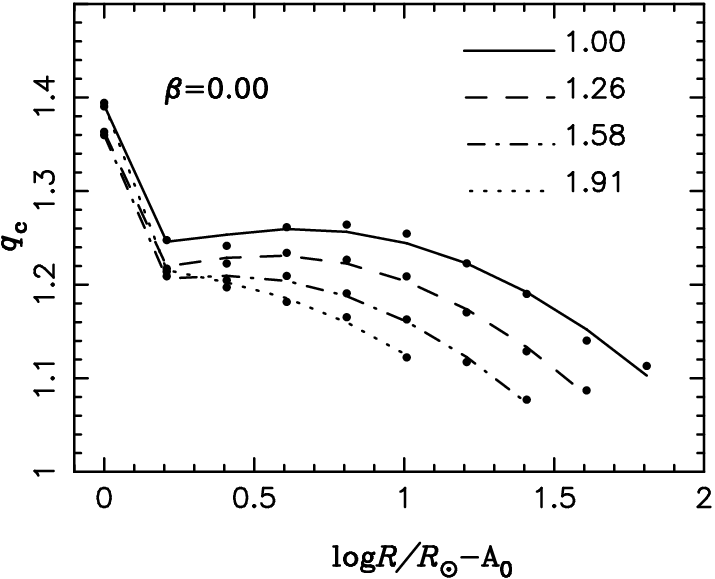}
 \caption{Critical mass ratio $q_{\rm c}$ for dynamically unstable
mass transfer. {\it Left panel:} a polytropic model. $f_{\rm c}$
is the core mass fraction of the mass donor. The cross, plus,
asterisk and circle lines are for mass transfer efficiency $\beta
=0.25$, 0.5, 0.75 and 1.0, respectively. The lost mass is assumed
to carry away the same specific angular momentum as pertains to
the mass donor. The solid line is from \cite{web88} for
conservative mass transfer (see also \cite{han01}). {\it Right
panel:} detailed binary evolution. It is for low-mass binaries
between a first-ascend giant-branch star and a main sequence star.
The giant star has a mass of 1.00, 1.26, 1.60 and $1.90M_\odot$.
The lines are from the fitting formulae of equations (4) to (6) in
\cite{chen08a}.}
   \label{qc}
\end{center}
\end{figure}

Before we describe the details of evolutionary channels to BSs
from binary evolution, we first introduce an important parameter
in binary evolution, the critical mass ratio, $q_{\rm c}$, for
dynamically unstable mass transfer, which is crucial to determine
the fate of a binary during mass transfer. The value of $q_{\rm
c}$ differs for different evolutionary stages of the primary at
the onset of mass transfer, and it has been well studied via
polytropic models (\cite{hje87,han01}) and detailed binary
evolutions (\cite{han02,chen08a}). Figure \ref{qc} shows two
examples on these studies. We may see obvious difference for
$q_{\rm c}$ between polytropic models and detailed binary
evolution, which in turn leads to differences on the products,
including BSs, after the mass transfer.

\subsection{Mass transfer}
In general, if the mass ratio $q=M_1/M_2$ (the donor/the accretor)
at the onset of mass transfer is lower than $q_{\rm c}$, mass
transfer is stable, and the accretor evolves upwards along the
main sequence in response to accretion, if it is a main-sequence
star, and becomes a blue straggler when it is more massive than
the turnoff of the host cluster.

This channel may produce binary BSs with various orbital periods
(\cite{chen04,chen08a}). If the primary is on the main sequence or
in the Hertzsprung gap at the onset of mass transfer, the products
are with short- or relatively short-orbital periods. These include
Algol systems which are still in a slow stage of mass transfer.
Mass transfer between giant stars and main-sequence companions may
produce long-orbital period BSs. The value of $q_{\rm c}$ here
significantly affects the BS number and orbital periods
(\cite{chen08a}).

Since the accreted material may be originating from the unclear
region of the mass donor, it is then rich in helium related to the
gainer's surface and has a higher mean molecular weight, which
results in secular instability during or after the accretion.
Thermohaline mixing will occur in this case. This mixing was once
believed to cause the surface abundance abnormality of the gainer
to be invisible. However, The study of \cite{chen04} showed no
distinction in surface composition between the models with and
without thermohaline mixing during mass transfer, although their
evolutionary tracks diverges. After the mass transfer, the C,N,O
abundance abnormalities may exists for about $10^8$ yr, comparable
to the lifetime of a typical BS. Thus, we could observe C,N,O
abundance abnormalities in BSs this way.

\subsection{Coalescence of contact binaries}
During mass transfer, the accretor likely fills its Roche Lobe and
the system becomes a contact binary. This contact binary
eventually coalesces as a single star (\cite{web76,egg00,lhz05}).
If both components are on the main sequence, their remnant is also
a main-sequence star and evolves similarly to a normal star with
that mass. So the remnant may be a BS if it is more massive than
the turnoff.

By assuming that the matter from the secondary homogeneously mixes
with the envelope of the primary and that no mass is lost from the
system during the merger process, \cite{chen08b} constructed some
mergers of contact binaries and studied their characteristics.
Their study shows that some mergers are on the left of the
zero-age main sequence defined by normal surface composition (i.e.
helium content Y = 0.28 with metallicity Z = 0.02 for Population
I) on a colour-magnitude diagram because of enhanced surface
helium content. In addition, the central hydrogen content of the
mergers is independent of mass. Thus, the concentration towards
the blue side of the main sequence with decreasing mass predicted
by {\cite{ss03}}, does not appear in their models. In fact, there
is no evidence for the concentration from observations.

In old clusters, angular momentum loss (AML) of low-mass binaries
induced by magnetic braking is a main factor to lead binaries to
be contact and merge finally. A simple estimation (\cite{chen08b})
shows that, in old clusters, BSs from the AML are much more
numerous than those from evolutionary effect only , indicating
that the AML of low-mass binaries makes a major contribution to
BSs in old cluster such as in NGC188, NGC 2682 etc.. In clusters
with intermediate age, e.g. in NGC 2660, the models of
\cite{chen08b} can account for several BSs. However, in the most
likely region on the colour-magnitude diagram, no BSs have been
observed yet. About 0.5 $M_\odot$ of mass loss in the merger
process is necessary to resolve this conflict.

\subsection{Coalescence due to dynamically unstable mass transfer}
If the mass ratio $q$ is larger than $q_{\rm c}$, mass transfer is
dynamically unstable, and a common envelope (CE) is formed. The CE
may be ejected if the orbital energy deposited in the envelope
overcomes its binding energy, or the binary will merge into a
single star. If the two components are main sequence stars, the
remnant of coalescence will be on the main sequence, and it is  a
BS if its mass is beyond the turn-off mass of the host cluster.

In this case, the core of the secondary spirals in quickly and
remains in the centre of the merger. The merger then has a
chemical composition similar to that of the primary, resembling
the result of smoothed particle hydrodynamic calculations
(\cite{lrs96,sill97,sill01}).

Binary coalescence of a contact binary or dynamically unstable
mass transfer is a popular hypothesis for single BSs
(\cite{mateo90,pol94,and06,chen08b}).

\section{Binary population synthesis}
We performed five sets of simulations for a Population I
composition (X = 0.70, Y = 0.28 and Z = 0.02) to systematically
investigate BS formation from primordial binary evolution. The age
ranges from 0.1 to 20 Gyr. The mass transfer efficiency, $\beta$,
which is defined as the mass fraction of the matter lost from the
primary accreted by the secondary, is an important parameter
remarkably affecting on the final results. Generally, we set
$\beta=1$ when the mass donor is on the main sequence and
$\beta=0.5$ otherwise. However this value is very unclear, and it
is likely higher than 0.5 when the mass donor is in HG while lower
than 0.5 when the mass donor is on FGB or AGB. So we also studied
the case of $\beta=1$ when the mass transfer begins in HG
(\cite{chen09}).

\subsection{Distribution on colour-magnitude diagrams}
\begin{figure}
\begin{center}
 \includegraphics[width=2.0in,angle=270]{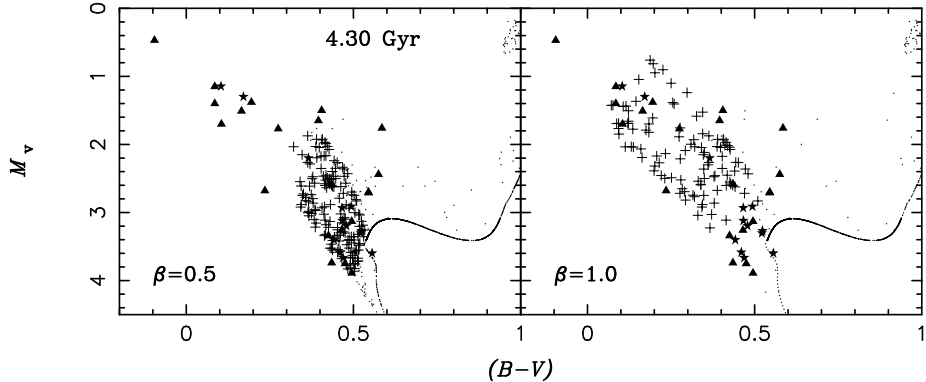}
 \caption{Colour-magnitude diagrams when the population is at
4.3Gyr. Here $\beta$ is the mass fraction of the matter lost from
the primary accreted by the secondary when the primary is in HG at
the onset of mass transfer. The crosses and stars are for BSs from
mass transfer and binary coalescence, respectively, and the small
dots are for other objects in the population. The triangles are
observed BSs of M67 from \cite{ss03}.}
   \label{cmd}
\end{center}
\end{figure}

Figure \ref{cmd} is colour-magnitude diagrams of a population of
4.3 Gyr for various $\beta$ values when the primary is in HG at
the onset of RLOF. From this figure we see that BSs from $\beta
=1$ may be more massive than those of $\beta =0.5$, indicating
that the high value of $\beta$ may produce BSs far away from the
turnoff. In particular, we may obtain BSs with masses larger than
2 times of the turnoff even these objects have very short
lifetimes. It is appropriate to assume a high value of $\beta$ for
old clusters, since the binaries contributing to BSs are less
massive than those in young clusters.

We also see in the figure that many BSs are below but bluer than
the turnoff. They may extend into the region 1 mag lower than the
turnoff. These objects are mainly from mass transfer and binary
coalescence of dynamically unstable mass transfer. They are less
evolved than the turnoff and may contribute more to the flux of V
band in the following evolutions. In general, most BSs from mass
transfer and binary coalescence of dynamically unstable mass
transfer are within 1.5 mag of the turnoff, while some from binary
coalescence of contact binaries may stay above the turnoff about
2.3 mag.

\subsection{The Specific Frequency }
The BS number from the simulations, $N_{\rm BS}$, depends slightly
on the population age in our simulations, while the specific
frequency, ${\rm log}F_{\rm BSS}$($ \equiv {\rm log}(N_{\rm
BS}/N_2)$, $N_{\rm 2}$ is the number of stars within 2 mag below
the main-sequence turnoff), heavily depends on the age due to the
increase of $N_{\rm 2}$. In all of the five sets, ${\rm log}F_{\rm
BSS}$ decreases with time first, and then increases when the age
is larger than 10 Gyr. The decrease of ${\rm log}F_{\rm BSS}$
before 1.5 Gyr comes from the increase of $N_{\rm 2}$.
Subsequently, the number of potential binaries which may
contribute to BSs decreases, leading to the formation of fewer
BSs. On the other hand, $N_2$ continues increasing. Thus, ${\rm
log}F_{\rm BSS}$ continues decreasing. Over time, the primaries in
long-orbital-period binaries gradually enter into AGB phase,
dramatically expand, and some of them may fill their Roche lobe
and start mass transfer. Due to large stellar winds in the AGB
phase, these mass donors at the onset of mass transfer are
probably much less massive than before and this mass transfer is
easily stabilized, resulting in some long-orbital-period BSs. As a
consequence, ${\rm log}F_{\rm BSS}$ begins to increase when the
age is longer than 10 Gyr. AML becomes more and more important for
BS formation with time and exceeds that from primordial binary
evolution when the age is older than 2.5 Gyr.

When we investigate the role of primordial binary evolution on BS
formation in Galactic open clusters, we found a serious problem
that the BS specific frequency obtained in our simulations is much
lower (only about 20 per cent of the observed) than that observed,
which may result from the following aspects. (1)Observational
errors, such as $N_2$ counted in clusters, the cluster ages and
the BS sample etc. For example, the study of \cite{car08} showed
that a large fraction of earlier BSs are actually field stars.
(2)The adopted Monte Carlo simulation parameters in the
simulation, including initial mass function, distributions of
initial ratio and initial separation. All these parameters in our
simulation are for field stars and it is very likely that these
parameters are different in clusters. For instance, recent studies
show that the initial mass function might be quite different
between disk stars and halo stars in the Galaxy (\cite{pol08}).
(3)Other channels e.g a more recent era of star formation and
dynamical interaction, to produce BSs in open clusters in addition
to primordial binary evolution and AML.

\subsection{Contribution to ISED}
\begin{figure}
\begin{center}
 \includegraphics[width=3.0in,angle=270]{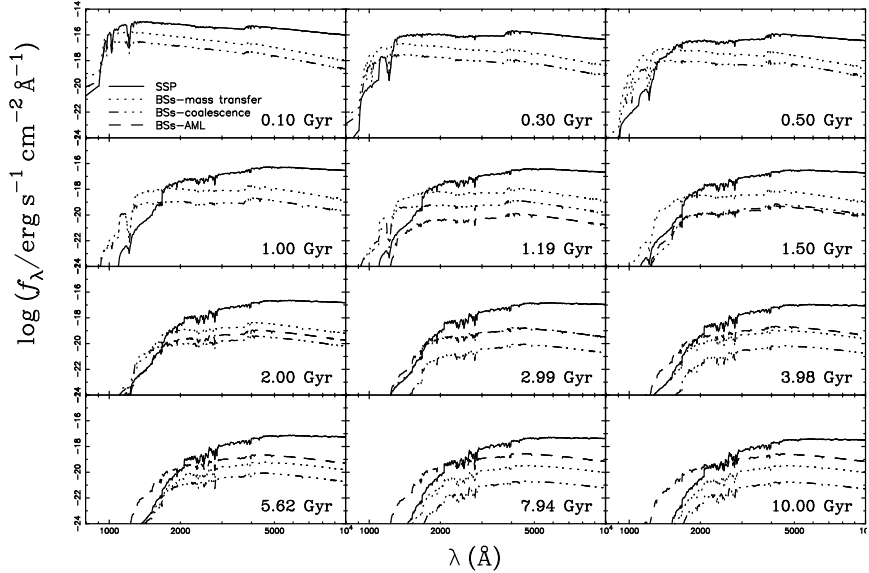}
 \caption{Integrated rest-frame intrinsic SEDs for a stellar
population with a mass of $1M_\odot$ at a distance of 10kpc with
$\beta =0.5$ when the primary is in HG at the onset of mass
transfer. The solid lines are for the results of SSP, which means
a population without binary interaction, and the others are the
contributions of BSs from different evolutionary channels. }
   \label{sed}
\end{center}
\end{figure}

We plotted ISEDs of a population for various cases in Fig.
\ref{sed}, where SSP means a population without binary
interaction. This figure shows that BSs resulting from binary
evolution are dominant contributors to the ISED in UV and blue
bands between 0.3 and 2.0 Gyr. The BSs and SSP have comparable
energy in UV and blue bands between 2 and 4 Gyr. The contribution
from AML becomes more important with time, and exceeds that from
primordial binary evolution for a population older than $\sim 3$
Gyr. Thus, primordial binaries are important contributors to BSs
over the whole age range.

Since BSs from a high $\beta$ have higher masses, their
contributions to ISED then become more important in ultraviolet
and blue bands (see Fig.5 in \cite{chen09}).

\subsection{Influences on the colours and ages}
Obviously, the excess spectral energy due to BSs in UV and blue
bands inevitably results in some changes in colours involving
these bands. Thus, we studied some Hubble colours and the results
are shown in Fig.4, from which we see that the colours are
affected from about 0.32 Gyr (all of the five colours shown in
this figure) to older than 10 Gyr (i.e F185W-F336W and
F218W-F336W), and the maximum difference may be up to 1.5 mag for
F170W-F336W.
\begin{figure}
\begin{center}
 \includegraphics[width=1.9in,angle=270]{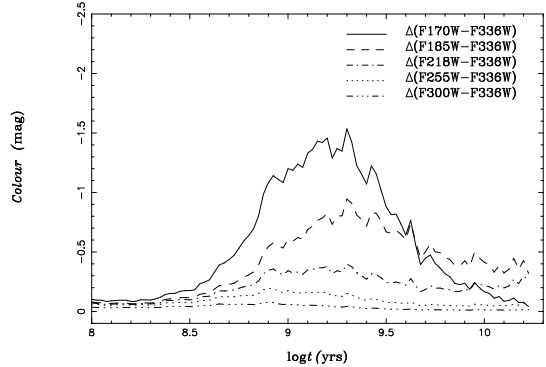}
 \caption{Colour differences versus age for a population with and without BSs.}
   \label{fig1}
\end{center}
\end{figure}

\section{Conclusions and Acknowledgments}
In this contribution, we summarized the linkage of binary
evolution and BSs, and BS characteristics. As well, we showed
binary population synthesis results of BSs from primordial binary
evolution, such as the distribution on CMD, the contribution to
ISED, the specific frequency and the influences on colours. This
work was in part supported by the Chinese National Science
Foundation (Grant Nos. 10603013 and 10973036,10821061 and
2007CB815406) and Yunnan National Science Foundation (Grant No.
08YJ041001).

\end{document}